\documentclass[a4paper,12pt]{article}
\usepackage[utf8]{inputenc}
\usepackage{multicol}
\title{On Some Statistical and Axiomatic Properties of
the Injury Severity Score
 }
\author{Nassim Dehouche\\
\small nassim.deh@mahidol.edu\\
\small Mahidol University International College\\
\small Salaya, 73170, Thailand.}
\usepackage[margin=1.2in]{geometry}
\date{}
\usepackage{multirow}
\usepackage{enumitem}
\usepackage{multirow}
\usepackage{colortbl}
\usepackage{graphicx}
\begin{document}

\maketitle

\begin{abstract}
The Injury Severity Score (ISS) is a standard aggregate indicator of the overall severity of multiple injuries to the human body. This score is calculated by summing the squares of the three highest values of the Abbreviated Injury Scale (AIS) grades across six body regions of a trauma victim. Despite its widespread usage over the past four decades, little is known in the (mostly medical) literature on the subject about the axiomatic and statistical properties of this quadratic aggregation score. To bridge this gap, the present paper studies the ISS from the perspective of recent advances in decision science. We demonstrate some statistical and axiomatic properties of the ISS as a multicrtieria aggregation procedure. Our study highlights some unintended, undesirable properties that stem from arbitrary choices in its design and  that call lead to bias in its use as a patient triage criterion.\\


\end{abstract}

\section{Introduction}
The Injury Severity Score (ISS) is a widely-used aggregation procedure for assessing injuries to multiple body parts and evaluating the emergency of care. The ISS aggregates multiple evaluations based on the Abbreviated Injury Scale (AIS) \cite{AIS}, a common evaluation scale for the severity of trauma to individual body parts. The AIS evaluates the severity of damage to each of nine body regions (head, face, neck, thorax, abdomen, spine, upper extremities, lower extremities, and external)  on a scale of 0 to 5\footnote{A grade of 6 additionally indicates untreatable injuries. This value being immaterial to the purpose of this paper, we will omit it from our analysis}. The ISS is the sum of the squares of the AIS scores of the three most severe injuries and is thus evaluated on a scale of $0$ to $75$. After reviewing past work on the ISS, and notably the seminal study \cite{ISS} that introduced this aggregation procedure, this paper questions the choice of a quadratic procedure relative to two other arbitrary aggregation functions (the sum and sum of cubes of the three highest AIS scores). Then we study some axiomatic properties of the ISS and propose that an injury severity aggregation procedure should be seen as an adjustment lever to optimize target criteria, rather than a rigid formula that seeks to capture fundamental aspects of the response of the human body to injuries with a quadratic formula (as has been wildly conjectured in the original study in the face of the high correlation of the ISS with mortality).

\section{Basic Definitions and Notation}

The ISS was introduced in a study by Baker et al. \cite{ISS}, which considered a sample of 2,128 motor vehicles occupants who were victims of accidents and admitted to one of 8 hospitals in the city of Baltimore, Maryland, USA, over a period of two years (1968-1969). For this sample, the study recorded a ratio of hospital admissions to deaths of 8:1. For individual hospitals, this ratio ranged from 5:1 to 60:1, indicating different levels of severity of injuries for the typical patient that each hospital received. Table \ref{1} presents the distribution of AIS for the main injury of each patient in the sample, while Table \ref{2} details the mortality rates corresponding to the highest AIS grade of patients. 

\begin{table}[!htbp]
\begin{center}
\resizebox{\columnwidth}{!}{
\begin{tabular}{|c|c|c|c|c|c|} 

\hline

\cellcolor[gray]{0.8} \textbf{AIS Grade} & \cellcolor[gray]{0.8} \textbf{Dead on arrival} & \cellcolor[gray]{0.8} \textbf{Dead later} & \cellcolor[gray]{0.8} \textbf{Survived} & \cellcolor[gray]{0.8} \textbf{Unknown} & \cellcolor[gray]{0.8} \textbf{Percentage}\\

\hline
1& 0 & 0 & 80 & 1 & 4\% \\
2& 0 & 2 & 437 & 1 & 20\% \\
3& 0 & 23 & 997 & 20 & 49\% \\
4& 0 & 30 & 229 & 3 & 13\% \\
5& 93 & 80 & 97 & 3 & 13\% \\
Unknown &1 & 0 & 12 & 0 & 1\% \\
\hline
\end{tabular}

}
\caption{\label{1}Distribution of AIS grades over the sample of 2,128 patients in \cite{ISS} }

\end{center}
\end{table}

The authors find that the ISS explains $49\%$ of the variance in mortality, in the study sample.

\begin{table}[!htbp]
\begin{center}

\begin{tabular}{|c|c|} 

\hline

\cellcolor[gray]{0.8} \textbf{Maximum AIS}&\cellcolor[gray]{0.8} \textbf{Percentage died} \\

\hline
1 &0\%  \\
2& .5\%\\
3& 3\%\\
4&16\%\\
5&64\%\\
\hline
\end{tabular}
\end{center}

\caption{\label{2}Mortality rates according to the three highest AIS grades for the sample of 2,128 patients in \cite{ISS} }

\end{table}

The AIS evaluates individual injuries to a body region as follows:
\begin{enumerate}[start=0]
\item No injury
\item Minor injury
\item Moderate injury
\item Serious injury
\item Severe injury
\item Critical injury
    
\end{enumerate}
To calculate the ISS the nine AIS body regions are grouped into
six: 
\begin{itemize}
\item $R_1:$ Head or neck
\item $R_2:$ Face
\item $R_3:$ Chest
\item $R_4:$ Abdominal or pelvic contents
\item $R_5:$ Extremities or pelvic girdle
\item $R_6:$ External
\end{itemize}

Formally, let us denote  $AIS=\{R_1, \dots, R_6\}$, the AIS scores of an injured patient over these six body regions, which we will also refer to as the patient's AIS profile. The computation of the ISS aggregates these score in two steps: 
\begin{enumerate}
\item The three highest AIS scores are determined, that is $A=\max(AIS)$, $B=\max(AIS-\{A\})$, and $C=\max(AIS-\{A,B\})$.
\item The sum of squares of $A$, $B$, and $C$ is calculated, that is $ISS=A^2+B^2+C^2$.
\end{enumerate}

\section{Statistical Properties}
\subsection{On the use of a quadratic aggregation function}
The first step of the ISS aggregation procedure (use of the three maxima) is justified in \cite{ISS} by the fact that considering the sum of squares of the AIS scores of the three most severe injuries considerably improved the correlation of the resulting score with mortality rates, when including the fourth highest AIS score had no appreciable effect. In this work, we will not analyze the first steps of the aggregation procedure and focus on the second. However, in section \ref{cardinal}, we show that statistical measures such as the correlation and standard deviation are not well suited for a variable such as the ISS, because they incorrectly assign it a cardinal value, which leads to inconsistent results. Thus, the criterion used to evaluate the appropriateness of the ISS may be flawed. We should also mention an existing variant to the first step of the aggregation procedure, that questions not the use of three maxima for the AIS but the choice of body regions over which they are calculated. A widely-used such variant has been introduced under the denomination New Injury Severity Score (NISS) \cite{NISS}. Instead, of considering the three most severely injured body regions, this variant considers the three most severe injuries overall, the reasoning being that the original ISS method can potentially disregard more severe injuries that happen to be in the same body region as the most severe injury. This medical modification is inconsequential to the analysis and claims made in this paper focusing on the intrinsic mathematical properties of the method. Our results apply to both variants.\\

Thus the main focus of this study is the second step of the aggregation procedure. Indeed, in \cite{ISS} the choice of aggregating the three maxima by summing their squares was rather lightly justified as "the simplest nonlinear function", without further explanations on the type of complexity being referred to. 
This justification will be put to question in the present work as the calculation of say the sum of cubes, or any other polynomial function of $A$, $B$, and $C$ is no more complex than that of the ISS. As for the use of linear functions (e.g. summing the three maxima), it is dismissed in similarly vague terms with the sentence "the quantitative relationship of the AIS scores is not known and is almost certainly nonlinear". The authors of the ISS further find that "the death rate for persons with two injuries of grades 4 and 3 was not comparable to that of persons with two injuries of grades 5 and 2 (sum = 7 in both cases)". 

\subsection{The scale of the ISS}

Table \ref{pos} describes the scales of the ISS ($A^2 + B^2 + C^2$), as well as the sum ($A + B + C$) and sum of cubes ($A^3 + B^3 + C^3$) functions. For $A,B,C \in \{0,1,2,3,4,5\}$, such that $A \geq B \geq C$ and excluding triplet $(0,0,0)$, there are $55$ possible $(A,B,C)$ triplets, resulting in $44$ distinct possible values of the ISS ($A^2 + B^2 + C^2$), as well as $13$ and $55$ distinct values of  ($A + B + C$) and ($A^3 + B^3 + C^3$), respectively.

\begin{table}[!htbp]

\begin{minipage}{.5\linewidth}
\resizebox{\columnwidth}{!}{
\begin{tabular}{|cccc|} 

\hline

\cellcolor[gray]{0.8} \textbf{Rank} & \cellcolor[gray]{0.8} $\mathbf{A + B + C}$   & \cellcolor[gray]{0.8} $\mathbf{A^2 + B^2 + C^2}$ & \cellcolor[gray]{0.8}$ \mathbf{A^3 + B^3 + C^3} $\\ 

\hline
1&	1&	1&	1\\
2&	2&	2&	2\\
3&	3&	3&	3\\
4&	4&	4&	8\\
5&	5&	5&	9\\
6&	6&	6&	10\\
7&	7&	8&	16\\
8&	8&	9&	17\\
9&	9&	10&	24\\
10&	10&	11&	27\\
11&	11&	12&	28\\
12&	12&	13&	29\\
13&	13&	14&	35\\
14&	14&	16&	36\\
15&	15&	17&	43\\
16& -		&	18&	54\\
17&	-	&	19&	55\\
18&	-	&	20&	62\\
19&	-	&	21&	64\\
20&	-	&	22&	65\\
21&	-	&	24&	66\\
22&	-	&	25&	72\\
23&	-	&	26&	73\\
24&	-	&	27&	80\\
25&	-	&	29&	81\\
26&	-	&	30&	91\\
27&	-	&	32&	92\\
28&	-	&	33&	99\\
\hline
\end{tabular}
}
\end{minipage}
\begin{minipage}{.5\linewidth}
\resizebox{\columnwidth}{!}{
\begin{tabular}{|cccc|} 

\hline

\cellcolor[gray]{0.8} \textbf{Rank} & \cellcolor[gray]{0.8} $\mathbf{A + B + C}$   &\cellcolor[gray]{0.8} $\mathbf{ A^2 + B^2 + C^2}$ & \cellcolor[gray]{0.8} $\mathbf{A^3 + B^3 + C^3} $\\ 
\hline
29&	-	&	34&	118\\
30&	-	&	35&	125\\
31&	-	&	36&	126\\
32&	-	&	38&	127\\
33&	-	&	41&	128\\
34&	-	&	42&	129\\
35&	-	&	43&	133\\
36&	-	&	45&	134\\
37&		-	&	48	&	136\\
38&		-	&	50&		141\\
39&		-	&	51&		152\\
40&		-	&	54&		153\\
41&		-	&	57&		155\\
42&		-	&	59&		160\\
43&		-	&	66&		179\\
44&		-	&	75&		189\\
45	&	-	&-	&	190\\
46	&-	&-	&		192\\
47	&-	&-	&		197\\
48	&-	&-&	216\\
49&-	&-	&			250\\
50&	-&-	&			251\\
51&	-&-	&			253\\
52&	-&-	&			258\\
53&	-&-	&			277\\
54&	-&-	&			314\\
55&	-&-	&			375\\
\hline
\end{tabular}
}
\end{minipage}

\caption{\label{pos} Possible grades and their rank, for the sum, sum of squares (ISS), and sum of cubes}

\end{table}

We have computed all rank reversals between the ISS, the sum, and the sum of cubes. In other words, the the number of pairs of injury profiles for which the rankings provided by the two aggregation functions are reversed. Among the $C_2^55=1485$ distinct, non-ordered pair of possible AIS profiles, we have identified the pairs for which there is discordance between $A^2 +B^2 +C^2$, $A^3 +B^3 +C^3$, and $A +B +C$, regarding the comparison of the pair. In other words, and for two patients $x$ and $y$, let $(A_x, B_x, C_x)$ and $(A_y, B_y, C_y)$ be their respective AIS profiles. We consider that there is discordance between the ISS and the sum of cubes aggregation function if ($A^2_x + B^2_x + C^2_x > A^2_y + B^2_y + C^2_y$ and $A^3_x + B^3_x + C^3_x < A^3_y + B^3_y + C^3_y$) or    ($A^2_x + B^2_x + C^2_x < A^2_y + B^2_y + C^2_y$ and $A^3_x + B^3_x + C^3_x > A^3_y + B^3_y + C^3_y$).  There exist $84$ pairs of profiles for which there is such a discordance, which represents $5.6\%$ of the $1485$ possible pairs of profiles (i.e. for a uniform distribution of AIS scores, the ISS and sum of cubes aggregation functions would disagree $5.6\%$ of the time). The ISS and the sum are in discordance for $8\%$ of the profiles, whereas the sum of cubes and the sum are in discordance for $14.81\%$ of the profile. Although a minority, these cases of discordance are non-neglectable, particularly for large volumes of patients.

\subsection{Association with mortality}

The seminal work \cite{ISS} relied on the data in Table \ref{yes}, which records the mortality rates for the AIS scores of the three most severe injuries, which we denote $A$, $B$ and $C$ by decreasing order of severity.

\begin{center}
\begin{figure}[!htbp]
\includegraphics[width=\columnwidth]{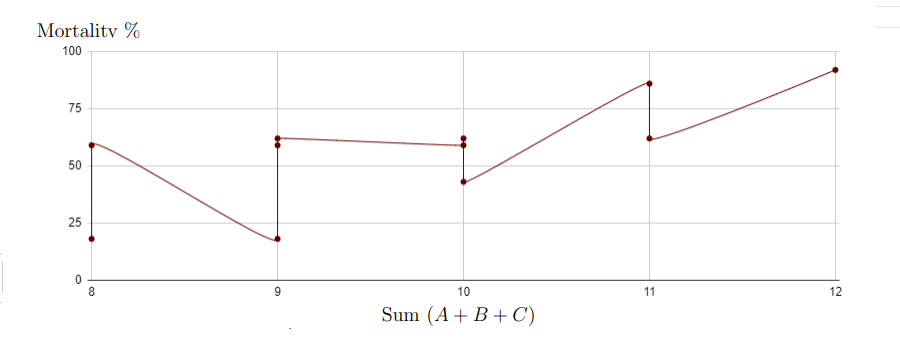}
\caption{\label{summor}Mortality rates according to sum of the three highest AIS scores for the sample of 2,128 patients in \cite{ISS} }
\end{figure}
\end{center}

\begin{center}
\begin{figure}[!htbp]

\includegraphics[width=\columnwidth]{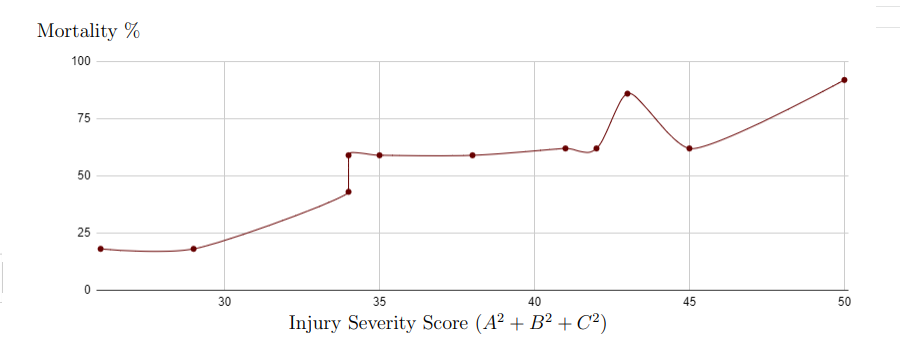}
\caption{\label{issmor}Mortality rates according to ISS (sum of squares of the three highest AIS scores) for the sample of 2,128 patients in \cite{ISS} }
\end{figure}
\end{center}

\begin{center}
\begin{figure}[!htbp]

\includegraphics[width=\columnwidth]{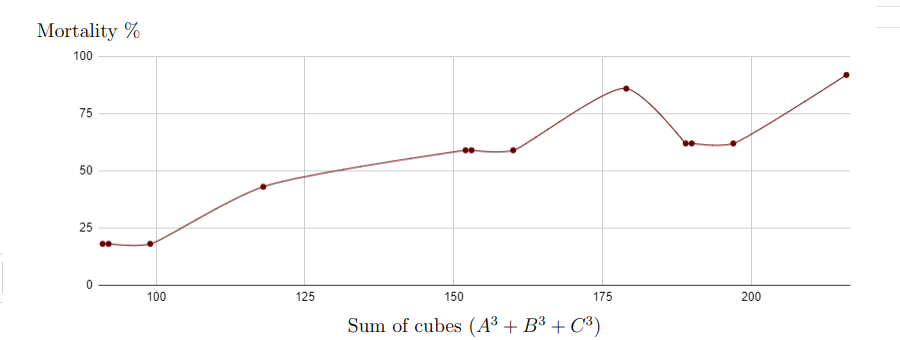}
\caption{\label{cubemor}Mortality rates according to sum of cubes of the three highest AIS scores for the sample of 2,128 patients in \cite{ISS} }
\end{figure}
\end{center}

\begin{table}[!htbp]
\begin{center}
\begin{tabular}{|c|cc|cc|cc|} 

\hline

 \textbf{Number of persons} & \multicolumn{2}{c}{102} & \multicolumn{2}{|c|}{78} & \multicolumn{2}{c|}{38}  \\

\hline

\textbf{Most severe injury $(A)$} & \multicolumn{2}{c}{4} & \multicolumn{2}{|c|}{5} & \multicolumn{2}{c|}{5}  \\

\hline
\textbf{Second most severe injury $(B)$} & \multicolumn{2}{c}{3} & \multicolumn{2}{|c|}{3} &\multicolumn{2}{c|}{4}  \\

\hline
 \textbf{Third most severe injury $(C)$} & 0-2 & 3 &0-2 &3 &0-2 & 3\\
\hline
\textbf{Percentage died} &18\%  &43\%  &59\% &86\% &62\% &92\% \\
\hline

\end{tabular}
\end{center}

\caption{\label{yes}Mortality by AIS scores of the three most severe injuries in \cite{ISS} }

\end{table}
The use of the ISS was supported in \cite{ISS} by the data reproduced in Table \ref{2}, in which we have additionally included the sums of the three most severe ISS, of their squares (the ISS), and of their cubes, and calculated the (Pearson product-moment) correlation and Mutual Information of each profile with mortality rates. Figure \ref{summor}, Figure \ref{issmor}, and Figure \ref{cubemor} respectively plot mortality rates according to sum, sum of squares (ISS), and sum of cubes of the three highest AIS scores for the sample of 2,128 patients in \cite{ISS}. 

\begin{table}[!htbp]
\begin{center}
\begin{tabular}{|c|c|c|c|c|c|c|} 

\hline

\cellcolor[gray]{0.8} $\mathbf{A}$ & \cellcolor[gray]{0.8} $\mathbf{B}$ & \cellcolor[gray]{0.8} $\mathbf{C}$ &\cellcolor[gray]{0.8} $\mathbf{A + B + C}$  & \cellcolor[gray]{0.8} $\mathbf{A^2 + B^2 + C^2}$  & \cellcolor[gray]{0.8} $\mathbf{A^3 + B^3 +C^3}$&\cellcolor[gray]{0.8} \textbf{Mortality}  \\
\cellcolor[gray]{0.8} &\cellcolor[gray]{0.8}   & \cellcolor[gray]{0.8}  &\cellcolor[gray]{0.8}  &  \cellcolor[gray]{0.8} \textbf{(ISS)}&\cellcolor[gray]{0.8}  &\cellcolor[gray]{0.8} \textbf{rate} \\
\hline
4&	3	&0	&7	&	25&91	&18\%\\
\hline
4&	3	&1	&8 &26		&92	&18\%\\
\hline
4&	3&	2& 9&	29&		99&	18\%\\
\hline
4&	3&	3& 10&	34&		118&	43\%\\
\hline
5&	3&	0&8& 	34&		152&	59\%\\
\hline
5&	3&	1&9&	35&		153&	59\%\\
\hline
5&	3&	2&10&	38&		160&	59\%\\
\hline
5&	3&	3&11&	43&179&	86\%\\
\hline
5&	4&	0& 9&	41&	189&	62\%\\
\hline
5&	4&	1&10&	42&		190&	62\%\\
\hline
5&	4&	2&11&	45&		197&	62\%\\
\hline
5&	4&	3&12&	50&		216&	92\%\\
\hline
\hline
\multicolumn{3}{|c|}{Correlation} & & & &\\
\multicolumn{3}{|c|}{with mortality} &0.77 & 0.92&0.92 &1.00\\
\hline
\multicolumn{3}{|c|}{Mutual Information} & & & &\\
\multicolumn{3}{|c|}{with mortality} &0.46 & 0.55&0.71 &1.00\\
\hline
\end{tabular}
\end{center}
\caption{\label{2}Mortality rates associated with the AIS profiles of the 2,128 patients in \cite{ISS} }

\end{table}

The high (Pearson's product-moment) correlation of the ISS and mortality has led \cite{ISS} to conjecture that this score "models a fundamental aspect of the human body's response to multiple injuries". Though it remains a practical heuristic for priority evaluation and patient triage, the initial promise of the ISS as an indicator of the mortality of multiple injuries, and the even more daring conjecture of \cite{ISS} that this quadratic function may capture fundamental properties of the response of human bodies to injuries have been tempered down by more mathematically rigorous, recent studies of the discrete possible values taken by the ISS. In \cite{MONO}, it has been found that mortality is non-monotonic with regards to the ISS, that is, mortality does not necessarily increases with successive values of ISS. 

Following the same reasoning as \cite{ISS}, we use correlation with mortality as a measure of the adequacy of the three aggregation procedures. The sum of the three highest AIS scores presents the lowest correlation with mortality with $77\%$ and Figure \ref{summor} conveniently illustrates the reason for the inadequacy of this aggregation procedure. As indicated by the number of vertical and horizontal segments in the graph, the sum, which only offers 15 possible distinct values represented in Table \ref{pos}, is not discriminant enough in relationship to mortality. However, the sum of squares (ISS) and the sum of cubes present similar levels of correlation with mortality rates, at $92\%$ and as Figure \ref{issmor}, and Figure \ref{cubemor} show that the ISS (with 44 distinct possible value versus 55 for the sum of cubes, cf. Table \ref{pos}) is less discriminant. All three functions are non-injective as evidenced by the existence of horizontal segments in the graphs. However, the relationship between the sum of cubes and mortality is of a functional nature (no vertical segments), as opposed to that of ISS with mortality. For instance, an ISS of $34$ corresponds to both mortality rates of $43\%$ and $59\%$. No such effects occur when considering the sum of cubes. However such undesirable effects cannot be evaluated by a coefficient of linear correlation, which would arbitrarily consider that the mortality rate associated with an ISS of $34$ is $52\%$, the average of $43\%$ and $59\%$.

\subsection{The AIS and ISS are not cardinal measures}

Measurement theory \cite{MEAS} assumes that there exist some empirical structure that one wishes to represent numerically (e.g. the body's response to multiple injuries) and defines strict qualitative properties that the empirical structure must verify in order to be represented numerically. Such numerical artifacts are said to possess an interval level of measurement if, throughout its scale, equal differences in the measure reflect equal differences in the empirical structure being measured. Nothing indicates that the AIS and even less so the ISS possess such a property. The AIS and ISS can be more modestly considered to possess an ordinal level of measurement, that is to say as indicators allowing the ranking of patients, e.g. for triage purposes. An ordinal measure is defined, by opposition to a cardinal one, as "a variable whose attributes can only be ranked" \cite{ordinal, AGRESTI}. For instance, we know that an underlying injury having an AIS score of $3$ is less severe than a $4$, which in turn is less severe than a $5$, but it remains unknown whether the distance between a $3$ and a $4$ is equal, greater, or smaller than the distance between a $4$ and a $5$. It is the practice of assigning the numerical values to the severity of these three injuries that sets the two numerical distances between them to be equal. The interpretation of the distances between ISS scores is similarly impossible. 
 Indeed, the consecutive values in the domain of the ISS, represented in Table \ref{pos} only reflect an increase in the severity of the overall injury (ordinal information), but the extent of that increase cannot be given any interpretation (it contains no cardinal information). For instance, $50, 51, 54$ are three consecutive values in the domain of the ISS, without any possible value between $51$ and $54$. A patient whose condition goes from an ISS of $50$ to $51$ and then from $51$ to $54$ would have seen the severity of their injury increase by two (ordinal) units, not four (cardinal) units.

Giving a cardinal meaning to the ISS could have been justified if the difference between two consecutive values of this scale kept increasing, reflecting a higher level of degradation as the severity of an injury increases, but this is not the case. In Table \ref{pos}, we can observe for instance that the gap between the thirty-second and thirty-third grades of the ISS (scores of $38$ and $41$, respectively) is wider than between the thirty-fourth and thirty-fifth grades (scores of $42$ and $43$, respectively).

\subsection{Pearson's correlation is not applicable to the ISS}
The value of the ISS is only ordinal, that is the information it provides is to rank the overall severity of injuries to multiple body regions of patients, and not measure any intrinsic property of these injuries. Further, \cite{SKEW} warns against considering the ISS/NISS as continuous statistical variables in correlation analyses with outcome measures (e.g. mortality), which has been the approach initially used to justify the quadratic aggregation of AIS grades in the original version of the ISS. If we accept the ISS as a purely ordinal indicator, a much simpler argument can be made to show that the very concept of measuring Person's correlation of the ISS with any other variable does not apply. Pearson's product-moment correlation is defined as the covariance of two variables divided by the product of their standard deviation \cite{PEARSON}. Focusing on the ISS, we can observe that the concept of standard deviation does not apply to this variable.

Consider the toy example in table \ref{var} in which we measure the standard deviation of ISS, in three samples of two patients each. The two patients in each sample are of two consecutive ranks, with regards to the ISS ($28th$ and $29th$, $32nd$ and $33rd$, as well as $34th$ and $35rd$ ranks, respectively). Note that the ISS profiles of a patient in consecutive samples only differs by one unit of AIS (e.g. the three samples could correspond to a similar degradation of patient 1 and of patient 2 injuries over three periods of time).

\begin{table}[!htbp]
\begin{center}
\begin{tabular}{|cccccc|} 
\hline
Sample & Patient 1   & Patient 2 & Patient 1 ISS& Patient 2 ISS& Variance of ISS\\
 & ISS profile  & ISS profile & (and rank) & (and rank)& in sample\\
\hline
A&(5,2,2) & (5,3,0)& 33 (28th) & 34 (29th)& 0.5\\
\hline
B&(5,3,2) & (5,4,0)&38 (32nd) &41 (33rd)&4.5\\
\hline
C&(5,3,3) & (5,4,1)&42 (34th)&43 (35th)&0.5\\
\hline
\end{tabular}

\caption{\label{var} The variance of the ISS arbitrarily increases because of the uneven gaps between consecutive grades of the ISS scale }
\end{center}
\end{table}

We observe a significantly higher standard deviation and thus variance in sample B than in sample A, which is not due to a wider dispersion of the severity of injuries in sample B, but is solely due to the cardinal properties of the ISS. There happens to be no possible ISS values between 38 and 41. The range of ISS goes back to one unit in sample C, and we find the same variance as in sample A. 

Thus, the very concept of a unit of deviation of the ISS is meaningless and no interpretation can be made of the standard deviation of this variable and hence of its covariance or Pearson correlation with any other variable. These concepts being based on that of a deviation of the observed ISS values relative to the mean, it is impossible to separate the amount of deviation that is due the observations and the amount due to the makings of ISS scale, with its uneven distances between grades.

The calculations of the standard deviation and variance of the ISS, as well as its Pearson's correlation with mortality and the analysis of said correlation does not account for the average and standard deviation of the distance between two consecutive Injury Severity Scores (they are not one and zero respectively). It implicitly consider this score to be cardinal (i.e. a measure of the amount of something).

 However for measures of mortality the average and standard deviation of the distance between two consecutive possible values are respectively one unit (depending on the decimal precision considered for mortality rates) and zero. 

\label{cardinal}

\subsection{Mutual Information as a more appropriate measure of the association between injury severity and mortality}
The ordinal nature of the ISS calls for the use of rank correlation. Spearman's rank-order correlation coefficient \cite{SPEARMAN, AGRESTI} could be more appropriate measurements of the association between ISS and mortality rates. Indeed, this statistic evaluates the monotonic association between two variables without utilizing ordinal information. However, it cannot be precisely evaluated in the presence of ties, which are common as seen in Figures \ref{summor}, \ref{issmor}, and \ref{cubemor}. Moreover, this indicator would be sensitive to the intrinsic variance of the ISS for consecutive values of the AIS, illustrated with the example in Table \ref{var}. A more robust measurement of the association between mortality and ISS would be offered by Mutual Information \cite{MI}. This more general indicator, which is less sensitive to the cardinal properties of random variables and is not limited to linear relationships, compares probability distributions as a whole and measures how different the joint probability distribution of two random variable is to the product of their marginal distributions. Thus Mutual Information $MI(X,Y)$, given by $MI(X,Y)=H(X)-H(X|Y)$ between two random variables $X$ and $Y$ is the amount of information (in bits) about one random variable that is gained by knowing the value of the other random variable, where $H(X)$ is the marginal entropy of $X$ and $H(X|Y)$ the conditional entropy of $X$ in regard to $Y$. In its normalized form, mutual information quantifies this amount of information relative to the intrinsic entropy of each random variable. The normalized mutual information $NMI(X,Y)$ between $X$ and $Y$ is thus given by $NMI(X,Y)=\frac{2\cdot MI(X,Y)}{H(X)+H(Y)}$. Using this indicator with the data in Table \ref{2}, for the three considered aggregation procedures and with p-values of order of magnitude $10^{-6}$, we find normalized amounts of Mutual Information of $0.46$, $0.55$, and $0.71$ between mortality rates in Table \ref{2} and the sum, sum of squares, and sum of cubes of AIS scores, respectively. For this data-set, there is thus a significantly higher amount of information concerning mortality rates contained in the sum of cubes than the sum of squares, which confirms and quantifies the visual insight gained from Figure \ref{issmor} and Figure \ref{cubemor} and suggests Mutual Information as a more appropriate measurement of the association between aggregate scores based on the AIS and mortality rates.

\section{Axiomatic properties}
For an AIS profile of the form $(A,B,C)$, we introduce the notation $[x_A,x_B,x_C]$  such that $-A \geq x_A \geq 6-A , -B \geq x_B \geq 6-B,$ and $-C \geq x_C \geq 6-C$ indicate a change in the AIS profile of a patient (i.e. an overall degradation or improvement of their injuries), resulting in a new AIS profile $(A+x_A,B+x_B,C+x_C)$. We assume, without loss of generality, that these changes maintain the the three most severe injuries located in the same three body regions (out of the six AIS body regions previously grouped). For instance, $[-1,0,+1]$ represents an improvement of the most severe injury of a patient by one AIS point (e.g. following care), and a degradation of their third most severe injury by one AIS point, without any change to their second most severe injury. These vectors can be conventionally added with ISS profiles to obtain the resulting ISS profiles, e.g. a patient whose ISS profile is $(4,3,2)$ would see their ISS profile become $(4,3,2)+[-1,0,1]=(3,3,3)$, following the above described change. Using this notation, we study some axiomatic properties \cite{Roy} of the ISS.

\subsection{Arbitrary compensation}
A multicriteria aggregation procedure is said to be compensatory if it allows for trade-offs between criteria, i.e. the possibility of compensating a disadvantage on some criteria by an advantage on other criteria \cite{Greco}. As such, the ISS being a simple aggregated score, it is a fully compensatory procedure, in that any disadvantage on any criterion (a lower AIS score) can be compensated by an advantage on any other criterion (a higher AIS score). For instance, improving the second most severe injury by one AIS point, while degrading the third most severe injury by two AIS points would bring the same change to the weighted sum, no matter its initial value.

Should a patient accept a medical procedure that improves your second most severe injury by one AIS point, but degrades your third most severe injury by two AIS points (for instance during transportation or waiting for said procedure)? Let us consider the toy example in Table \ref{trade}.\\

\begin{table}[!htbp]
\begin{center}
\begin{tabular}{|c|cc|} 

\hline
\cellcolor[gray]{0.8} \textbf{Patient} & \cellcolor[gray]{0.8} \textbf{Patient 1}   & \cellcolor[gray]{0.8} \textbf{Patient 2} \\
Initial ISS Profile &$(5,4,3)$ & $(4,4,4)$\\
Initial ISS &$50$ &$48$\\
\hline
Change &$[0,+1,-2]$ &$[0,+1,-2]$\\
\hline
Resulting ISS Profile &$(5,5,1)$ &$(5,4,2)$\\
Resulting ISS &$51$ & $45$\\
\hline
\end{tabular}

\caption{\label{trade} An improvement for Patient 2 (decrease in ISS) is a degradation for Patient 1 (increase in ISS)}
\end{center}
\end{table}

An improvement in Patient 2's condition (decrease in ISS) is a degradation in Patient 1's condition (increase in ISS).

This property of the ISS function is arbitrary. It does not have anything to do with the fact that Patient 1 was initially in a slightly worse state than Patient 2. It is due to the fact that trade-offs between AIS scores $A$, $b$ and $C$ in the calculation of the ISS do not obey a fixed compensation rate. The very notion of improvement or degradation of the AIS score is thus meaningless. A weighted sum wouldn't suffer from this inconsistency, as the trade-off rates between criteria would be constant and defined by their weights.

\subsection{Arbitrary rank reversals for identical changes}
Table \ref{antimonot} shows a toy example in which Patient 1 and Patient 2 receive twice the same procedure (an improvement of their most severe injury by one AIS point followed by an improvement of their second most severe injury by one AIS point). Initially, the overall condition of Patient 2 (ISS of $33$) is worse than that of Patient 1 (ISS of $32$). However, after the first procedure the order of severity of the conditions of the two patients alternates to Patient 1 (ISS of $25$) being worse off than patient 2 (ISS of $24$) and then back to Patient 2 (ISS of $21$) being in a worse condition than Patient 1 (ISS of $20$), after the second procedure. Moreover, Table \ref{monot} shows a similar alternation of priority but with the condition of the two patients progressively degrading over time). In a situation where the ISS is used as a triage rule, the order of priority between the two patients would arbitrarily alternate, although the degradation of their states would be identical.

\begin{table}[!htbp]
\begin{center}
\begin{tabular}{|c|cc|}

\hline
\cellcolor[gray]{0.8} \textbf{Patient} & \cellcolor[gray]{0.8} \textbf{Patient 1}   & \cellcolor[gray]{0.8} \textbf{Patient 2} \\

\hline
Initial ISS Profile &$(4,4,0)$ &$(5,2,2)$\\
Initial ISS &$32$ & $33$\\
\hline
Change &$[-1,0,0]$ &$[-1,0,0]$\\
\hline
Resulting ISS Profile &$(4,3,0)$ &$(4,2,2)$\\
Resulting ISS &$25$ & $24$\\
\hline
Change &$[0,-1,0]$ &$[0,-1,0]$\\
\hline

Resulting ISS Profile &$(4,2,0)$ & $(4,2,1)$\\
Resulting ISS &$20$ &$21$\\
\hline
\end{tabular}

\caption{\label{antimonot} The order of priority of the two patients arbitrarily alternates despite an identical improvement of one of their AIS grades}
\end{center}
\end{table}

\begin{table}[!htbp]
\begin{center}
\begin{tabular}{|c|cc|}

\hline
\cellcolor[gray]{0.8} \textbf{Patient} & \cellcolor[gray]{0.8} \textbf{Patient 1}   & \cellcolor[gray]{0.8} \textbf{Patient 2} \\

\hline
Initial ISS Profile &$(4,4,0)$ &$(5,2,2)$\\
Initial ISS &$32$ & $33$\\
\hline
Change &$[0,+1,0]$ &$[0,+1,0]$\\
\hline
Resulting ISS Profile &$(5,4,0)$ &$(5,3,2)$\\
Resulting ISS &$41$ & $38$\\
\hline
Change &$[0,0,+1]$ &$[0,0,+1]$\\
\hline

Resulting ISS Profile &$(5,4,1)$ & $(5,3,3)$\\
Resulting ISS &$42$ &$43$\\
\hline
\end{tabular}

\caption{\label{monot} The order of priority of the two patients arbitrarily alternates despite an identical degradation of one of their AIS grades}
\end{center}
\end{table}

\subsection{Independence}
 The independence property states that identical performance on one or more criteria should not influence the way two alternatives compare \cite{ROY}. A transformation that maintains the value of the criterion equal should not change the way alternatives compare. 
In Table \ref{ex}, we consider two pairs of ISS profiles, Patient 1 and Patient 2 versus Patient 3 and Patient 4. The only difference between these two pairs concerns the AIS score of the most severe injury ($3$ and $4$ for patient 1 and patient 2 respectively, $4$ and $5$ for patient 3 and patient 4 respectively). An identical change, $[0,+1,0]$. is applied twice to the second most severe; it gains one point of severity. The two pairs of patients show an identical level of severity,  in their second and third most severe injuries before and after the transformation, respectively $(.,2,0)$ and $(.,0,0)$. However, the change leads to two different outcomes. Patient 1 condition ($ISS=13$), which was initially less severe than that of Patient 2 ($ISS=16$), becomes more severe ($18>17$), whereas the order of priority of Patient 3 and Patient 4 remain unchanged ($20<25$ and $25<26$).

\begin{table}[!htbp]
\begin{center}
\begin{tabular}{|c|cc||cc|}

\hline
\cellcolor[gray]{0.8} \textbf{Patient} & \cellcolor[gray]{0.8} \textbf{Patient 1} & \cellcolor[gray]{0.8} \textbf{Patient 2} & \cellcolor[gray]{0.8} \textbf{Patient 3} & \cellcolor[gray]{0.8} \textbf{Patient 4}\\

\hline
Initial ISS Profile &$(3,\mathbf{2},\mathbf{0})$ &$(4,\mathbf{0},\mathbf{0})$ &$(4,\mathbf{2},\mathbf{0})$ &$(5,\mathbf{0},\mathbf{0})$\\
Initial ISS &$13$ & $16$ & $20$ & $25$\\
\hline
Change &$[0,+1,0]$ &$[0,+1,0]$&$[0,+1,0]$&$[0,+1,0]$\\
\hline
Resulting ISS Profile &$(3,\mathbf{3},\mathbf{0})$ &$(4,\mathbf{1},\mathbf{0})$ &$(4,\mathbf{3},\mathbf{0})$ &$(5,\mathbf{1},\mathbf{0})$\\
Resulting ISS &$18$ & $17$& $25$ & $26$\\
\hline

\end{tabular}

\caption{\label{ex} An identical change to the second most severe injury \textit{ceteris paribus} leads to different outcomes}
\end{center}
\end{table}

\section{Conclusion}

The choice of an aggregation function is a highly sensible one that impacts mortality rates. The ISS is neither optimal in terms of its correlation with mortality nor in regards to its axiomatic properties. By attempting to be two things at the same time (a cardinal measurement of the body's response to multiple injuries as well as an ordinal triage rule), the Injury Severity Score and similar, competing indices (New Injury Severity Score, Exponential Severity Score \cite{EISS}, etc.) achieve sub-optimal results in both regards. A complex, fundamental property such as the physiological response to injury is unlikely to be universally captured by a simple mathematical function (the ISS) of ordinal mathematical measures (the AIS). If it existed such a function would be unlikely to be stumbled upon with arguments such as "the simplest nonlinear relationship is quadratic". However, research has mostly gone in this direction and proposals compete on which function offers the highest correlation with mortality. We have shown such measurements (as well as those of the standard deviation/variance of the ISS) to be unfounded.
We recommend system-thinking rather than setting an arbitrary formula in stone and conjecturing that it captures an essential property of the physiological response to injuries, when the sum of cubes, another arbitrary formula, shows better association with mortality. The choice of an aggregation function to be used for AIS scores (ISS, sum of cubes, or any other function) should be made on a case by case basis, through simulations for the specific distribution of AIS scores in an emergency department, in a way that optimizes target criteria. In our view, the unfounded, classical view in the literature of the ISS as a cardinal measurement of the severity of injuries and the ensuing correlation analyses with mortality have somehow hindered this actionable line of research.

\end{document}